\pdfoutput=1
\documentclass[pre,aps,twocolumn,showpacs,floatfix,nofootinbib,superscriptaddress]{revtex4-1}
\usepackage{subfigure}
\usepackage{amssymb}
\usepackage{amsfonts}
\usepackage{amsmath}
\usepackage{amsthm}
\usepackage{epsfig}
\usepackage{graphicx}
\usepackage{cancel}
\usepackage{xcolor}
\usepackage{color}
\usepackage[latin1]{inputenc}
\usepackage[hidelinks,unicode=true]{hyperref}
\usepackage{comment}
\usepackage[normalem]{ulem}
\hypersetup{colorlinks=true,
 	linkcolor=blue,
 	urlcolor=blue,
 	citecolor=blue,
 	pdfhighlight=/N
 }
\renewcommand{\vec}{\mathbf}

\begin{document}
\title{A geometrical interpretation of critical exponents} 
\author{Henrique A. Lima}
\email{Henrique\_adl@hotmail.com}
\address{International Center of Physics, Institute of Physics, University of Brasilia, 70910-900, Brasilia, Federal District, Brazil}
\author{Edwin E. Mozo Luis}
\email{emozo@id.uff.br}
\address{Instituto de F\'{\i}sica, Universidade Federal Fluminense, Avenida Litor\^{a}nea s/n, 24210-340, Niter\'{o}i, RJ, Brazil}
\author{Ismael S. S. Carrasco}
\email{ismael.carrasco@unb.br}
\address{International Center of Physics, Institute of Physics, University of Brasilia, 70910-900, Brasilia, Federal District, Brazil}
\author{Alex Hansen}
\email{alex.hansen@ntnu.no}
\address{PoreLab, Department of Physics, Norwegian University of Science and Technology, NO-7491 Trondheim, Norway}
\author{Fernando A. Oliveira}
\email{faooliveira@gmail.com}
\address{International Center of Physics, Institute of Physics, University of Brasilia, 70910-900, Brasilia, Federal District, Brazil}
\address{Instituto de F\'{\i}sica, Universidade Federal Fluminense, Avenida Litor\^{a}nea s/n, 24210-340, Niter\'{o}i, RJ, Brazil}
\begin{abstract}
We develop a hypothesis that the dynamics of equilibrium systems at criticality have their dynamics constricted to a fractal subspace. We relate the correlation fractal dimension associated with this subspace to the Fisher critical exponent controlling the singularity associated with the correlation function.  This fractal subspace is different from that which is associated with the order parameter.  We propose a relation between the correlation fractal dimension and the order parameter fractal dimension.  The fractal subspace we identify has as a defining property that  the correlation function is restored at the critical point by restricting the
dynamics this way. We determine the correlation fractal dimension of the 2d Ising model and validate it by computer simulations. We discuss growth models briefly in this context.  

\end{abstract}
\maketitle
\label{Int}
{\bf Introduction -} Symmetry and spontaneous symmetry breaking  are fundamental concepts for understanding Nature, in particular, to investigate the phases of matter. Phase transitions are typically described through symmetry breaking, which appears normally in the order parameter. We shall discuss here an important type of ``geometrical transition" which may appear in many natural processes, such as growth and second order phase transitions. Magnetic materials are excellent physical systems to look for symmetry or symmetry  breaking. The Ising Hamiltonian system exhibits the symmetric paramagnetic phase, and the symmetry-breaking phases (ferro and antiferromagnetic). In addition to the ordinary phases, it is possible to  construct additional fractal symmetry-protected topological (FSPT) phases via a decorated defect approach (see Ref.\ \cite{Devakul19} and references {therein}).

Our goal is to propose and develop a geometric interpretation of the critical exponents describing the singularities that appear in equilibrium or out of equilibrium systems with many degrees of freedom.  The idea is that the  {\it spatial\/} activity of such systems near or at singularities is restricted to a fractal subset of space, characterized by a fractal dimension $d_f$.  This allows us to interpret the critical exponents describing the singularity in terms of this fractal dimension.  We note that this is different from the notion of strange attractors which appear in phase space \cite{Strogatz00}.  We develop this idea in the context of the Ising model and the Kardar-Parisi-Zhang growth equation in the following --- one being example of an equilibrium system, the other a system out of equilibrium. 

{\bf Models -} The Ising model Hamiltonian is 
\begin{equation}
    H=- J \sum_{(i,j)} \sigma_i\cdot\sigma_j-h_0\sum_i \sigma_i,
\end{equation}
where the double sum $(i,j)$ runs over nearest neighboring spins. $J$ is the spin-spin coupling constant and $h_0$ is an external magnetic field, which in this work we shall take as zero. The spins $\sigma_i$ take on values $\pm 1$. We define the fluctuation $\psi(\vec{i})=\sigma_i-\langle \sigma_i\rangle$, where  $\langle\cdots \rangle$ means ensemble average. We define as well  the correlation function $G(r)=\langle \psi(\vec{r}+\vec{i})\psi(\vec{i})\rangle$, { whose Fourier transform in the small-$k$ limit is $\tilde{G}(k)=(k^2+\kappa^2)^{-1}$ which in real space yields $G(r)=r^{2-d} \exp(-r/\rho)$  where $\rho= \kappa^{-1}$ is the correlation length~\cite{Cardy96}. Note that $G(r)$ obeys the equation}
\begin{equation}
(-\nabla^2 +\kappa^2)G(r)=\delta^d(r).
\label{G}
\end{equation}
{This mean field solution breaks down at the transition, necessitating an empirical correction through the introduction of Fisher exponent~\cite{Fisher64} $\eta$,}
\begin{equation}
\label{G2}
G(r) \propto
\begin{cases}
r^{2-d} \exp(-r/\rho) , &\text{ if~~ } r>\rho,\\
r^{2-d-\eta}, &\text{ if~~ }  r \ll\rho.\\
\end{cases}
\end{equation}
 { Our main objective in the following is to interpret the small $r$ behavior in the context of fractal geometry.}
 
For $T$ close to the critical temperature $T_c$, the  correlation length diverges as
\begin{equation}
\label{rhodivergence}
\rho \propto |T-T_c|^{-\nu}.
\end{equation}
  The correlation length critical exponent $\nu$ is related to the critical specific heat exponent $\alpha$ through the hyperscaling relation~\cite{Cardy96}
\begin{equation}
\label{alpha}
\alpha=2-d\nu,
\end{equation}
thus associating a thermodynamic variable with the divergence of the correlation length.  We have as well the Fisher scaling relation~\cite{Fisher64}
\begin{equation}
\label{gamma}
\gamma =(2-\eta)\nu.
\end{equation}

{\bf Fractal dimension -}  {Fractals are statistically self-similar geometric objects characterized by non-integer dimensionalities \cite{feder2013fractals}. Interpreting the scaling behavior seen at critical points
as manifestations of fractal geometries is a well-researched field \cite{Suzuki8, Kroger00,Devakul19}.  The relation between the order parameter exponent $\beta$ and
the fractal dimension of the ordered phase $d_l$ was first proposed by Suzuki \cite{Suzuki8}
\begin{equation}
\label{dl}
d_l=d-\frac{\beta}{\nu}\;.
\end{equation}
If the critical point is the percolation one, the fractal structure is the infinite percolating cluster at the transition \cite{Grimmett06,Cruz23}.  In general, one associates it with the largest ordered cluster at the critical point \cite{Kroger00}. The fractal dimension predicted by Eq.\ (\ref{dl}), $d_l=15/8$ for the 2d Ising model,
coincides with that of Coniglio for the two state Potts model based on a mapping to the two-dimensional Coulomb gas \cite{Coniglio89}.  Cambier and Nauenberg  \cite{Cambier86} find a numerical value $d_l=1.90$. On the other hand,
Stella and Vanderzande \cite{Stella89} use conformal field theory to predict $d_l=187/96$, a slightly larger value than predicted by Eq.\ (\ref{dl}). 
}

{The question we pose here is whether the Fisher exponent $\eta$, defined in Eq.\ (\ref{G2}), may be interpreted in the context of the fractal structure characterized by the fractal dimension in Eq.\ (\ref{dl}).  Intuitively, such a relation
should exist, as $\eta$ describes the decay of the correlation function reflecting
the spatial structure of the ordered phase at criticality.}  

{{\bf Fractal dynamics-} Motivated by recent progress on response functions in growth phenomena 
based on assuming an underlying fractal dynamics \cite{GomesFilho21, GomesFilho21b,Anjos21,Luis22,GomesFilho24}, we reconstruct the steps that lead us to Eq.\ (\ref{G2}). Thus, as  $T \rightarrow T_c$ we modify Eq.\ (\ref{G}) to reflect that the dynamics are primarily confined to the fractal space of the spin clusters.}  

{We do so by taking as a starting point that the fractal structure which lies behind the dynamics is {\it not\/} the same as the fractal structure generated by the order parameter. They are characterized by different fractal dimensions, $d_l$ and
$d_f$. We will in the following refer to them as the {\it order parameter fractal
dimension\/} and the {\it correlation fractal dimension\/} respectively. Further on, we
will establish a relation between the two fractal dimensions, see Eq.\ (\ref{dl2}).}  

Muslih and Agrawal \cite{Muslih10,Muslih10b} obtained the Riesz fractional derivative of order $\zeta$, $0<\zeta<1$, associated  with fractal dimension $d_f$, $d-1<d_f<d$, as 
\begin{equation}
\label{SG3}
(-\nabla^2)^\zeta \left(\frac{1}{|\vec{r}-\vec{r'}|^{d_f-2\zeta}} \right) = \frac{4^{\zeta} \pi^{d_f/2} \Gamma(\zeta)}{ \Gamma(d_f/2-\zeta)} \delta^{d_f}(\vec{r}-\vec{r'}),
\end{equation}
where the fractional $\delta$-function satisfies
\begin{equation}
\label{DD1}
 \int \delta^{d_f}(\vec{r}-\vec{r'})f({ \vec{r'}})d^{d_f}\vec r'=f({\vec{r}}), 
\end{equation}
for any continuous function $f({\vec{r}})$. The simplest way to justify Eq.\ (\ref{SG3}) is to note that the left
and the right hand side of the equation have to scale in the same way when we scale $|{\bf r}| \to \lambda |{\bf r}|$.
The modified Laplacian was introduced by Laskin~\cite{Laskin07} to obtain a fractional Schr{\"o}dinger equation and consequently a fractional quantum mechanics. Eq.\ (\ref{SG3}) was formulated as a fractional version of the Poisson equation.
{Taking into account that at the transition  $\kappa =1/\rho \rightarrow 0$ at the critical point, we replace Eq.\ (\ref{G}) by}
\begin{equation}
\label{G3}
(-\nabla^2)^\zeta G(r)=\delta^{d_f}(r),
\end{equation}
{to reflect the dynamics being restricted to a dynamics fractal structure characterized by the correlation fractal dimension $d_f$. We have set the proportionality constant equal to one.}

We compare Eq.\ (\ref{SG3}) to Eq.\ (\ref{G3}) to obtain $G(r) \propto r^{2\zeta-d_f}$. We demand that the correlation function obtained with Eq.\ (\ref{SG3}) behaves like the one obtained with Eq.\ (\ref{G2}), yielding
\begin{equation}
\label{dfs}
    [d-d_f]-2[1-\zeta]+\eta=0.
\end{equation}

Returning to the work of Muslih and Agrawal \cite{Muslih10,Muslih10b}, we may proceed to construct a fractional version of Gauss' law
\begin{equation}
    \label{gauss}
    \oint_{\partial V} d^{d-1}\vec r\ {\vec n }\cdot {\bf E} = Q,
\end{equation}
where $\bf E$ is the field, $\vec n$ is a vector normal to the surface $\partial V$, and $Q$ is the total charge within the volume $V$.
We assume a sphere of radius $r$ containing a fractal distribution of charges. Hence,
\begin{equation}
\label{gauss2}
Q\propto r^{d_l}\;.
\end{equation}
The field in the radial direction is given by the Riesz fractional derivative
\begin{equation}
\label{gauss3}
E_r=-\frac{\partial^\zeta \Theta}{\partial r^\zeta}\;,
\end{equation}
where $\Theta$ is the potential from Eq.\ (\ref{SG3}),
\begin{equation}
\label{gauss3.5}
\Theta({\bf r})=\int \frac{g({\bf r}')\ d^{d_l}\vec r'}{|{\bf r}-{\bf r}'|^{d_f-2\zeta}},
\end{equation}
where $g(\vec r)$ is the charge density.
Rescaling ${\bf r}\to \lambda {\bf r}$ in Eq.\ (\ref{gauss}) then leads to the relation $\lambda^{d+d_l-d_f-1+\zeta}=\lambda^{d_l}$, 
which gives
\begin{equation}
\label{gauss4}
d-d_f=1-\zeta.
\end{equation}
This relation must also be true for the correlation length problem, as mathematically they are the same.  Hence, combining this equation with Eq.\ (\ref{dfs}), we find
\begin{equation}
\label{dff}
 \eta=d-d_f.
\end{equation}
Thus, the Fisher exponent in the correlation function $G(r)$, represents the deviation {of the correlation fractal dimension} from the integer dimension. Both $\zeta$ and $d_f$ are within the previously defined limits $0<\zeta<1$ and $d-1 \leq d_f \leq d$.

\begin{table}[h!]
       	\begin{tabular}{c|ccc}
		\hline	\hline
\hspace{5mm}	$d$\hspace{5mm}  & \hspace{5mm} $2$ \hspace{5mm}   & \hspace{5mm} $3$ \hspace{5mm} & \hspace{5mm} $4$ \hspace{5mm}   \\ \hline
{	$\beta$} & $1/8 $   &  $ 0.3265(3) $ &  $ 1/2$  \\ 
{		$\nu$} & $1$   &     $ 0.6300(3) $    & $1/2$  \\ 
{  		$d_l$} & $15/8$   &     $ 2.4817(5) $    & $3$  \\ 
		$\eta$ & $1/4$   &     $ 0.0364(5) $    & $0$  \\ 
		$d_f$ & $7/4$   &     $ 2.9636(5)$  &  $4$ \\
        $\zeta$ & $3/4$   &     $ 0.9636(5)$  &  $1$ \\
		 \hline
		 	\hline
	\end{tabular}	
	\caption{  Values of critical exponents for  the Ising universality class for $d=2$, 3 and 4 dimensions. $d_l$ from~\cite{Coniglio89},  $d_f$ from Eq. (\ref{dl2}) while $\eta$ from
    Eq. (\ref{dff}). Note the symmetric deviation for both the fractal dimension $d_f=d-\eta$ and the fractional derivative $\zeta=1-\eta$. For $d=3$ we use the results  of Pelissetto and Vicari~\cite{Pelissetto02}. }
   \label{Table1}
\end{table}

In order to establish a connection between the order parameter fractal dimension $d_l$ and the correlation fractal dimension $d_f$, we combine Eqs.\ (\ref{alpha}), (\ref{gamma}), (\ref{dl}) and (\ref{dff}) with the  Rushbrooke equality\footnote[1]{ Although Rushbrooke equality has been largely confirmed in the equilibrium phase transition, it has recently been questioned by Fytas {\bf et al} for the random-field Ising model~\cite{Fytas16}. If it breaks,  Eq. (\ref{dl2}) would also break down, but not our main result Eq. (\ref{dff})}. 
\begin{equation}
\label{rush}
\alpha +2\beta +\gamma=2
\end{equation}
to obtain
\begin{equation}
\label{dl2}
    d_f=2(d_l-1).
\end{equation}
In Table \ref{Table1}, we present the values of $d_l$ and $d_f$ for the Ising model in 2, 3, and 4 dimensions, alongside the corresponding critical exponents. We note that the equation is fulfilled for all dimensions shown. From this, $d_f$ and $\eta$ can be independently determined. Furthermore, it is noteworthy that as $d$ increases, $d_f$ approaches $d$, reaching its value at the upper critical dimension  $d=d_c=d_f=4$. 

\begin{figure}[htbp]
	\centering
	\includegraphics[width=0.8\columnwidth]{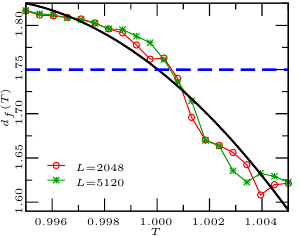}
        \caption{The {dynamic} fractal properties of the $2d$ Ising model. We exhibit  the fractal dimension $d_f(T)$ as a function of temperature $T$ for a squared lattice with  $L=2048$, and $L=5120$. The horizontal dashed blue line is {the} theoretical value $d_f=1.75$ at $T=T_c$ . The continuous black line  is the  the  function $f(T)=7/4+a(T-T_c)+b(T-T_c)^2$, after adjusting to the data } 
	\label{figfractal}
\end{figure}

 {\bf Monte Carlo tests on the Ising model-} To obtain numerical values for $d_f$, we use a specific box-counting method appropriate for this situation (see Supplementary Material \cite{SM}).

In Fig.\ \ref{figfractal} we exhibit $d_f(T)$ as function of $T$.  We use a square  lattice and the same procedure as {Fig.} $S3$. The plots exhibit the data for  $L=2048$ and $L=5120$. We observe large fluctuations for $T>1$. As observed in the supplementary material, fluctuations above $T_c$ are larger than those below. Furthermore, aside from fluctuations, the simulations suggest a universal behavior independent of $L$ for the sizes utilized here. Performing finite size scaling for each temperature around $T_c$ would demand significant computational power. Also, it takes a considerable amount of time to obtain  $d_f(T)$ as we do not use the CUDA process for the fractal program.  Thus we use only {a} few points around $T_c$. As expected  the fluctuations are larger in the region close to $T_c$, and they decrease as we increase the number of realizations (see table I supplementary material). For $T \approx T_c$,  $d_f(T) \approx f(T)=7/4+a(T-T_c)+b(T-T_c)^2$, which we use to determine $T_c=0.99996 \pm 0.00025$ for $L=2048$ and $T_c=1.00007 \pm 0.0003$ for $L=5120$.  Values for $L=1024 \times m$, $m=1,2,..6$ are reported in Table I. The results support {with good precision} our theoretical value $d_f=7/4$. Unfortunately, the numerical method is not good for three dimensions. 


{\bf Fluctuation-Dissipation Theorem-} There is a relation between the underlying assumptions behind this work and the Fluctuation-Dissipation Theorem (FDT). Correlation functions represent a simple form of fluctuation-dissipation relation, see the review~\cite{Gomesfilho23}, which are easily associated with the susceptibility~\cite{Goldenfeld18}.  
 {We claim that the violation of the FDT as in Eq.\ (\ref{G2}), is due to the change of geometry from an Euclidean to a fractal. Violation of the FDT is well-known in the literature for structural glasses~\cite{Grigera99,Ricci-Tersenghi00,Crisanti03,Barrat98,Bellon02,Bellon06}, random-exchange Heisenberg chain~\cite{Vainstein05}, proteins~\cite{Hayashi07}, KPZ dynamics~\cite{Kardar86,Rodriguez19}, mesoscopic radioactive heat transfer ~\cite{Perez-Madrid09,Averin10}, and ballistic diffusion as well \cite{Costa03,Costa06,Lapas07,Lapas08}. The FDT breaks down when there is a breaking of ergodicity \cite{GomesFilho21,Costa03,Wen23}. The FTD fails for critical points in the sense that it cannot explain Eq.\ (\ref{G2}) for $ r \ll\rho$, where the $\eta$ exponent appears as a need for consistency with the exponents equalities, e.g., the Fisher scaling relation Eq.\ (\ref{gamma}).   In fact, the response function Eq.\ (\ref{G}) in the Euclidean formulation yields $\eta=0$ in disagreement with reality, thus displaying a failure of the FDT. Moving  to a fractal formulation, i.e., replacing Eq.\ (\ref{G}) by Eq.\ (\ref{G3}), and thus recovering Eq.\ (\ref{G2}) {with a non-zero} $\eta=d-d_f$, Eq.\ (\ref{dff}) appears directly, without any extra assumptions. The FDT is restored in this way.}


{\bf Growth dynamics-} {We now turn to a brief description of  growth phenomena in the context of these results.}

Two quantities play an important role in growth, the average height, $ \langle h(t) \rangle $, and the standard deviation, which is also known as the roughness or the surface width, 
$w(L,t)= \left[ \langle h^2(t) \rangle - \langle h(t) \rangle^2\right]^{1/2}$.
Here the average is taken over space. The roughness is a very important physical quantity since many physical phenomena are controlled by it~\cite{Edwards82,Kardar86,Barabasi95,Gomes19,Reis05,Rodrigues15,Alves16,Carrasco18,Krug92,Krug97,Derrida98,Meakin86,Daryaei20,Hansen00}. For many growth processes, the roughness, $w(L,t)$, increases with time until reaches a saturated  {value} $w_s$, i.e., $w(t \rightarrow \infty)=w_s$. The time evolution of the roughness is well described by Family-Vicsek scaling~\cite{Barabasi95,Rodrigues24},  
\begin{equation}
\label{Sc1}
w(L,t)=
\begin{cases}
ct^{\chi/z} , &\text{ if~~ } t \ll t_\times,\\
w_s \propto L^\chi, &\text{ if~~ } t \gg t_\times,\\
\end{cases}
\end{equation}
with $t_{\times} \propto L^z$. Here $\chi$ and $z$ are the roughness and dynamic exponents, respectively.

{The surface roughness exponent $\chi$ may be related to an interface fractal dimension $d_i$ through the expression
\begin{equation}
\label{boxes}
d_i=
\begin{cases}
2-\chi , &\text{ if~~ } d=1,\ 2\;, \\
d-\chi, &\text{ if~~ } d \geq 2\;,\\
\end{cases}
\end{equation}
via box counting \cite{feder2013fractals,Barabasi95}.}

{Gomez-Filho et al.\ \cite{GomesFilho24} have recently demonstrated the relation between the roughness exponent $\chi$ and therefore the interface fractal dimension $d_i$, and the FDT, finding that it is also
restored in this system when the noise is restricted to a fractal which 
dimensionality is related to $d_i$.}


{\bf Conclusion -} We have introduced a new fractal structure 
associated with the dynamic correlations at equilibrium critical points, which 
is described by a correlation fractal dimension, which we relate to the Fisher exponent $\eta$ through Eq.\ (\ref{dff}). We have done this by substituting the correlation function Eq.\ (\ref{G}) by an equation based on the Riesz fractional derivative, Eq.\ (\ref{G3}) at the critical point. This equation defines the fractal subspace defined by the correlations.  By this substitution the correlation function is restored at the critical point.

{We obtain a correlation fractal dimension $d_f=7/4$ for the 2d Ising model from the critical exponents listed in Table \ref{Table1}.  We support this result by direct numerical simulations.}

{Lastly, we have briefly discuss growth phenomena in the context of the present work by pointing to recent work by Gomez-Filho et al.\ \cite{GomesFilho24}.}
{We note that scaling yields $N-2$ relations, such as  Eqs.\ (\ref{alpha}) and (\ref{gamma}), for the $N$ critical exponents.  Relations (\ref{dl}) and  (\ref{dff})  give us two  new equations, but they also introduce two new variables $d_l$ and $d_f$. In addition, we have Eq.\ (\ref{dl2}) between $d_f$ and $d_l$. As this relation is derived from the $N-2$ relations between the critical exponents we already have, so it does not constitute an independent relation.  On the other hand, if we could construct a theory to obtain both fractal dimensions, we would also obtain all the exponents associated with the critical point.}

Furer exploration of this connection, in conjunction with the present findings, could precede a novel approach to the physics of phase transitions, especially in dynamic phase transitions~\cite{Santos24} and disordered systems~\cite{Vainstein05}.

{\bf Acknowledgments -} This work was supported by 
the Conselho Nacional de Desenvolvimento Cient\'{i}fico e Tecnol\'{o}gico (CNPq), Grant No.  CNPq-01300.008811/2022-51 and  Funda\c{c}\~ao de Apoio a Pesquisa do Rio de Janeiro (FAPERJ), Grant No. SEI-260003/005791/2022 (E.E.M.L), and Funda\c{c}\~ao de Apoio a Pesquisa do Rio de Janeiro (FAPERJ), Grant No. E-26/203953/2022 (F.A.O.), and Funda\c{c}\~ao de Apoio a Pesquisa do Distrito Federal (FAPDF), Grant No.\ 00193-00001817/2023-43.  It was also partly supported in part by the Research Council of Norway through its Centres of Excellence funding scheme, Project Number 262644 (A.H.). 

\bibliography{referencesV1}

\clearpage

\appendix


\onecolumngrid

\section*{Supplemental material}

\section{Monte Carlo Study of the Ising Model}
In this supplemental material~\cite{Lima24} we investigate the Ising model in a two-dimensional lattice of size $L\times L$ along the lines described in the main article. The Ising model is described by the Hamiltonian
\begin{equation}
	H=- J \sum_i^{L^2} \sum_k \sigma_i\sigma_{i+k},
\end{equation}
where $J$ is the {coupling constant}, $i$ ranges over the lattice and $k$ over the nearest neighbor.  There is no external magnetic field present. In order to obtain the thermodynamical properties of the model, we use the checkerboard Metropolis algorithm~\cite{Romero_2020, yang2019high}.
We focus on the magnetization $m(T,t)$, which depends on the absolute temperature $T$ and evolves with the number of Monte Carlo time steps $t$ (in units of $L^2$ spin flips), being defined as
\begin{equation}
	\label{m}
	m(T,t)=\langle \sigma \rangle =\frac{1}{L^2} \sum_i^{L^2}  \sigma_i.
\end{equation}
To parallelize the algorithm we use the Compute Unified Device Architecture (CUDA)~\cite{sitecuda1, sitecuda2, sitecuda3}. In the following, we calibrate the results from the algorithm against already known results in order to judge the accuracy of our results before delving into the calculations pertinent to the main article. 

Our first step is to observe the evolution of $m(T,t)$  as a function of time $t$. We use the ensemble average Eq.\ (\ref{m}) with lateral size $L = 1024$ and fixed temperature $T$. We see that after time $t_i$ the magnetization reaches a stable average value 
\begin{equation}
	\label{mT}
	M(T)=\frac{1}{t_f-t_i+1}\sum_{t=t_i}^{t_f} m(T,t),
\end{equation}
while fluctuating around it. Here $t_f$ is the maximum running time.  Thus the final average $M(T)$ is taken over $N_r=L^2(t_f-t_i+1)$ spin configurations.

\begin{figure}[hbt!]
	\centering
	\includegraphics[width=11.2cm]{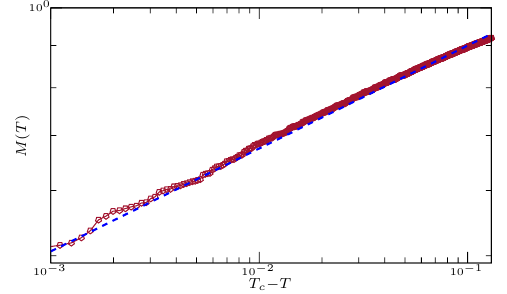}
	\caption{The magnetization  $M(T)$  as a function of  $T_c-T$, in log scale. $M(T)$ is the time average of  $m(T,t)$.
		The blue dashed straight line  is the result of a fit of  $M(T) \propto (T_c-T)^\beta $ to the data. From the fit we get $T_c=1.00056(7)$ and $\beta=0.125(1)$, { which are close to the exact values $T_c=1$ and} $\beta=1/8$.}
	\label{fig:S2}
\end{figure}

In Figure \ref{fig:S2} we show  $M(T)$  as a function of temperature $T_c-T$ in log scale. Here $L=1024$, $t_i=1000$
and $t_f=2500$, as defined earlier, time is taken in units of $L^2$ flips.  As expected for small $T$, $M(T) \approx 1$, while for $T \approx T_c$, $M(T)$ drops to zero. This way we only display the region close to $T_c$ for which  $M(T) \propto (T_c-T)^\beta $. From the fit to the data (blue  traced line), we obtain $T_c=1.00056(7)$ and $\beta=0.125(1)$ close to the exact value $\beta=1/8$.

In order to obtain the correlation length $\rho$ and the Fisher exponent $\eta$, we use 
the correlation function
{
	\begin{equation}
		\label{G0}
		G(r)=\frac{1}{t_f-t_i+1}\sum_{t=t_i}^{t_f}\left[\frac{1}{ML^2}\sum_{i=1}^{L^2}\sum_{j=1}^M\psi_i\psi_j\right],
	\end{equation}
	where $\psi_i =\sigma_i-m$ is the fluctuation the order parameter, Eq. (\ref{m}), in the site $i$, which runs over the lattice, while $j$ run over the $M$ sites with a distance $r$ of $i$.  The expression within the brackets corresponds to the correlation function for a given temperature at a Monte Carlo step $t$, where $t_i$ is after the thermalization is reached. To improve precision, we average the results over multiple Monte Carlo steps to reduce fluctuations. Finally, for each temperature, we fit the data to the function
	\begin{equation}
		\label{G}
		G(r) =c\frac{\exp(-r/\rho)}{r^{\eta}},
	\end{equation}
	where $c$ is a constant. This equation combines the possibility of displaying power laws and exponential.} Note that for $\rho\gg r$ we  get Eq. (3) of the main text with $d=2$. For a lattice with $L=1024$ we use $1 \leq r \leq 20$,  in this way, only $\rho > 20$  can be considered.

\begin{figure}[hbt!]
	\centering
	\includegraphics[width=10.2cm]{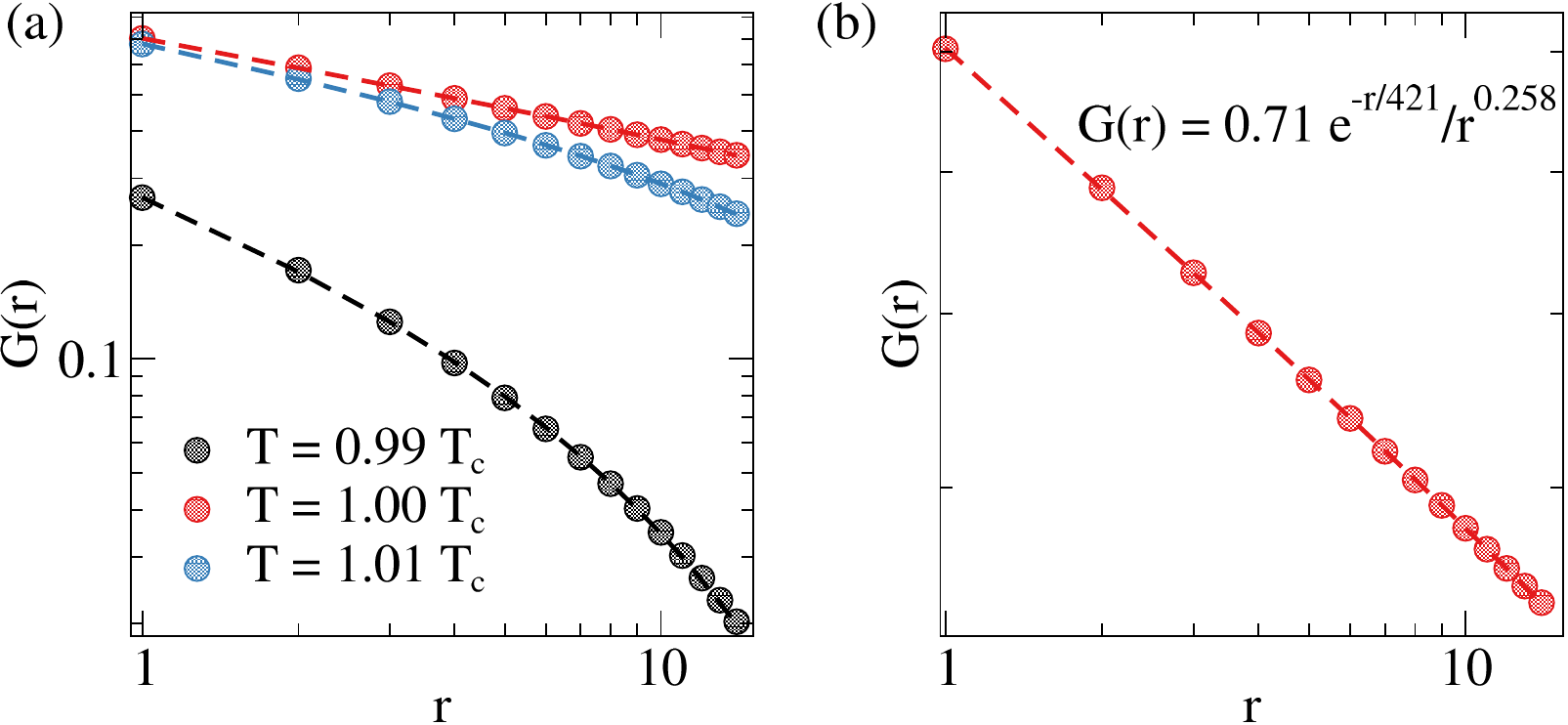}
	\caption{ {(a) The correlation function  $G (r)$  as a function of distance $r$ for the specified temperatures. The dashed lines represent the results of the fit using Eq. (\ref{G}). (b) Zoom in the result for $T=T_c$, highlighting the agreement between the data and the fit, where the parameters $\rho = 421(3)$ and $\eta = 0.258(1)$ are obtained.}}
	\label{fig:G}
\end{figure}
\begin{figure}[hbt!]
	\centering
	\includegraphics[width=11.2cm]{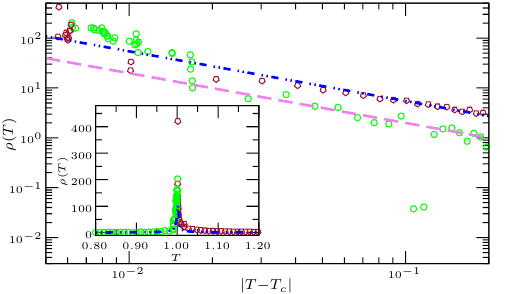}
	\caption{The correlation length  $\rho (T)$  as a function of temperature $T$. Here $L=1024$. In the inset we observe the singularity at $T_c$ which is the best indication of a phase transition. The points are from the simulations and the traced line is the adjustable function Eq.\ (\ref{rho}), we exhibit the points for $T <T_c$ and $T>T_c$.  From that we obtain  $T_c=1.00105(4)$ and $\nu=1.004(4) $.}
	\label{fig:S3}
\end{figure}
\begin{figure}[t]
	\begin{center}  	  \includegraphics[ height=6cm]{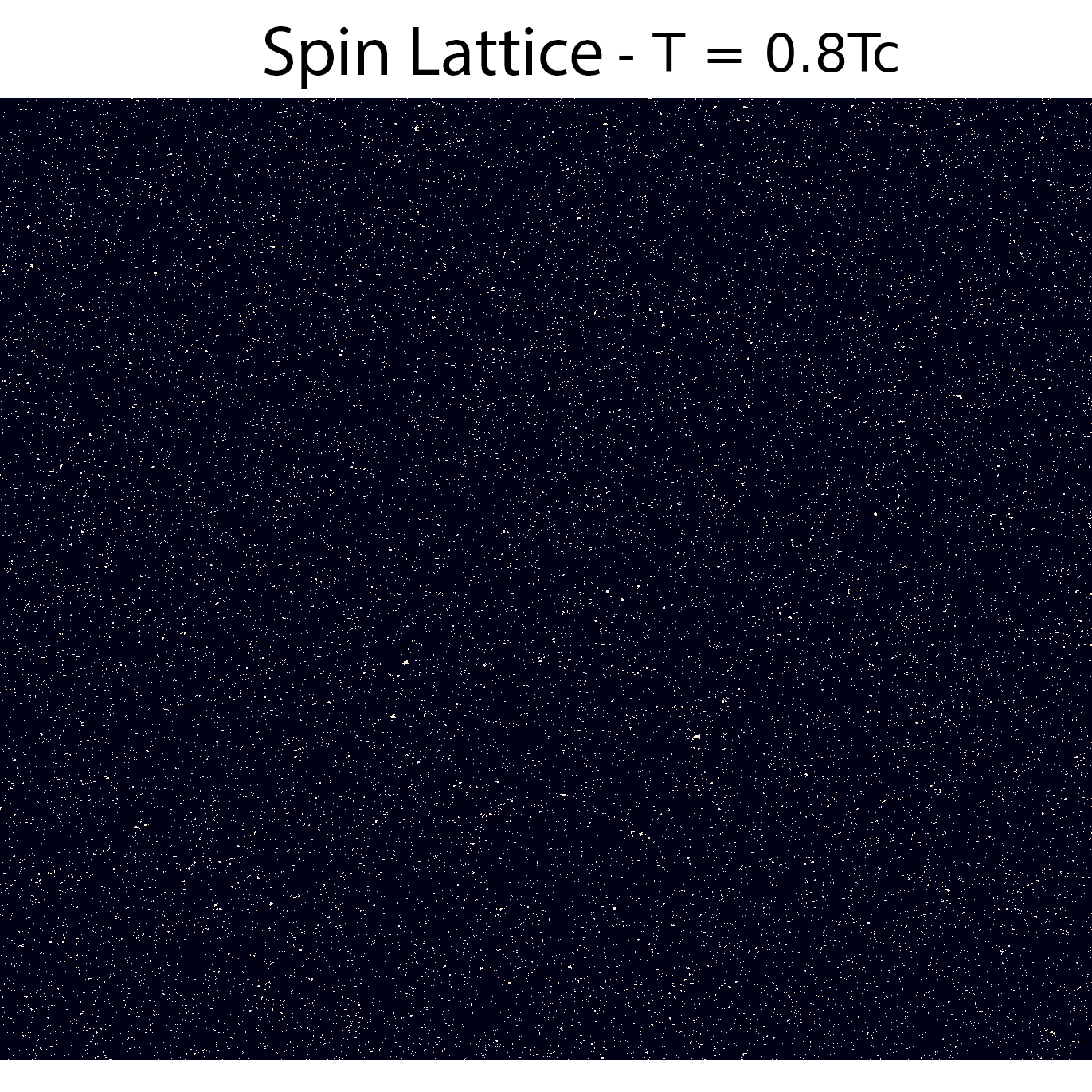}
		\includegraphics[height=6cm]{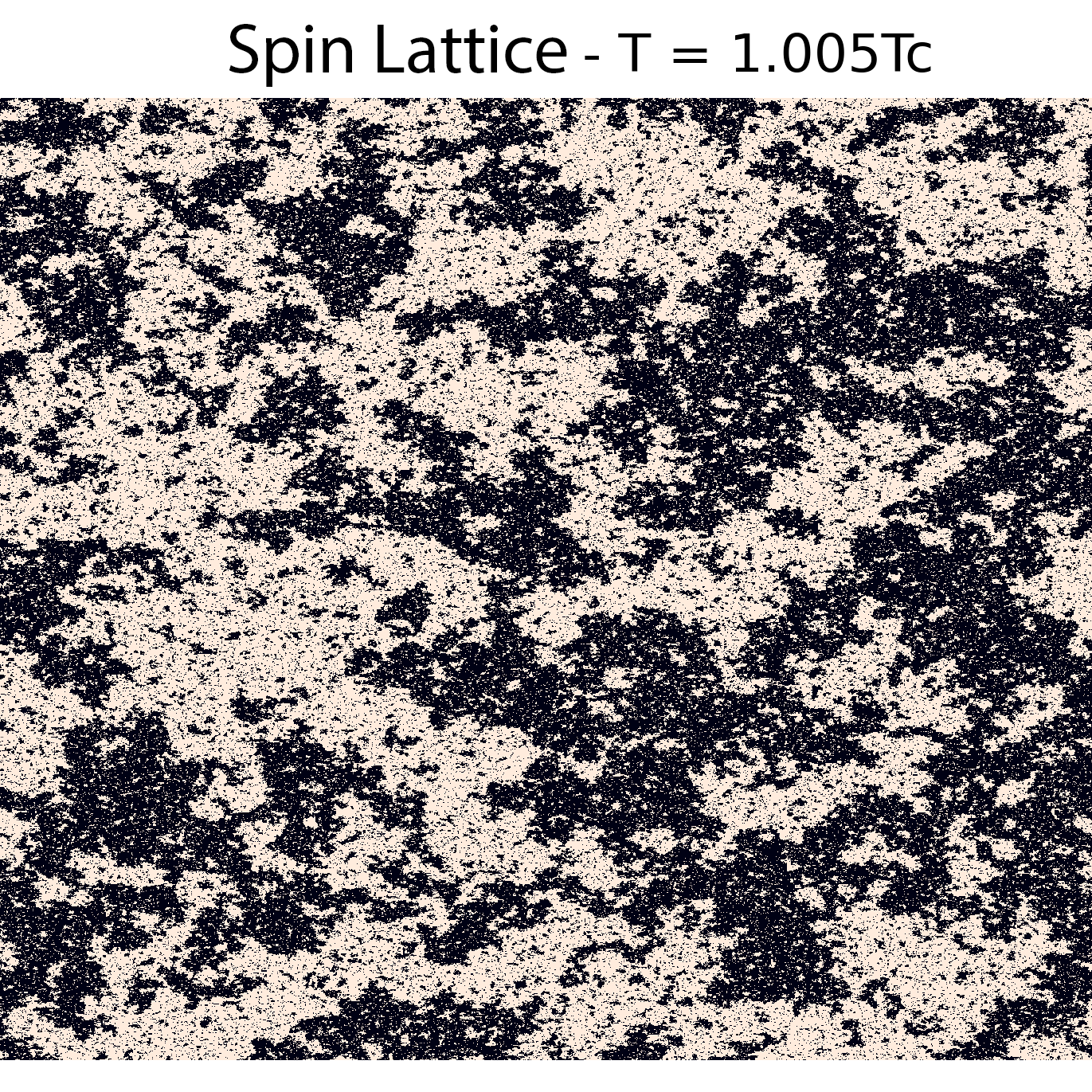}
		\caption{\label{Geom} Spin Matrix (a) $T=0.8T_c$, in this homogeneous medium $ |\nabla^2G(r)| << | \kappa^2 G(r)|$ ; (b) $T=1.005T_c$, heterogeneous medium $ |\nabla^2G(r)| >>  | \kappa^2 G(r)|$, clusters require fractal analysis .}
	\end{center}
\end{figure}

\begin{figure}[htbp]
	\centering
	\includegraphics[width=9.2cm]{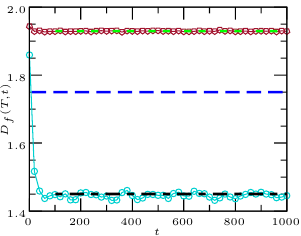}
	\caption{The instantaneous fractal dimension $D_f(T,t)$ of the $2d$ Ising model as a function of the number of Monte Carlo steps per site $t$ for a fixed temperature $T$. After the thermalization, the fractal dimension oscillates around the equilibrium value $d_f(T)$. The middle dashed {blue} line is the expected value $d_f=7/4$ at the transition. Upper curve $T=0.95$  and $d_f(0.95)=1.937(2)$ {(dashed green line)}; Lower curve $T=1.15$  and $d_f(1.15)=1.45(3)$ {(dashed black line)}. The fluctuations above $T_c$ are larger than those below $T_c$.}
	\label{figS4}
\end{figure}

{Figure \ref{fig:G} presents the correlation function $G(r)$ as a function of $r$ for  temperatures $T=0.99 T_c$, $T=1.0014 T_c \approx T_c$, and $T=1.10 T_c$. The dashed lines represent the fit of Eq. (\ref{G}). To provide a clearer view of the behavior at the critical temperature, in panel (b) we show again the results for $T=T_c$. In this case, the fit yields $\rho = 421(3)$ and $\eta = 0.258(1)$. The large value of $\rho$ indicates a long-range correlation, where $\exp(-r/\rho) \rightarrow 1$, leading to a power-law behavior $G(r) \sim r^{-\eta}$.}

In Figure \ref{fig:S3} we show the correlation length $\rho$ versus $T$. We see the expected divergence behavior near $T_c$, where 
\begin{equation}
	\label{rho}
	\rho(T) \propto |T-T_c|^{-\nu}.
\end{equation}
Obviously, for a finite system, $\rho$ is always limited by the system size $L$. From the fit\footnote[1]{ We adjust the function $\rho(T)=c_1+b_1|T-T_c|^{-\nu}$ to each side of the curve.} we  obtain  $T_c=1.00105(4)$ and $\nu=1.004(4) $ again very close to the exact values. 

\section{Calculation of the fractal dimension of the Ising model at criticality using the box-counting method}

{With our procedure for obtaining thermodynamic quantities near the transition now established, we can focus on our primary objective: determining the fractal dimension (as defined in the main text) for the two-dimensional Ising model. {This was inspired by recent results from growth models~\cite{Anjos21,GomesFilho21,Luis22,Luis23,GomesFilho24}.}  It is important to note that in homogeneous regions of the system $|\nabla^2G(r)| << | \kappa^2 G(r)|$, whereas at the edges of the spin cluster domains $ |\nabla^2G(r)| >>  | \kappa^2 G(r)|$. To illustrate this, Fig. \ref{Geom} displays the spin matrix, with white representing up spins and black representing down spins, to show how the Laplacian behaves both near and far from the transition. In particular: (a) Far from the transition, $\nabla^2G(s)\rightarrow0$; (b) Near the transition, $\kappa=\rho^{-1}\rightarrow 0$, $\kappa^2$ contribution becomes less relevant and $(-\nabla^2)G(r) \rightarrow  (-\nabla^2)^\zeta G(r)$. Thus, it is important to note that the main contribution of the $\nabla^2$ arises at the edges of the clusters, where fluctuations are most pronounced. Therefore, the fractal object relevant to the correlation function (Eq. 2 in the main text) is composed of the edges of all clusters. To measure the fractal dimension $d_f$, we can apply the traditional box-counting method used to estimate the hull fractal dimension of percolation clusters at $T_c$, but applied to all clusters rather than restricting to the percolating one.}

{The algorithm involves covering the system with a grid of squares of varying sizes. For $T<T_c$, we consider spins aligned in the same direction as the magnetization, while for $T \geq T_c$ spins in both directions yield the same result. We then count the number of boxes, as a function of box size, that contain any spin with the defined orientation. Finally, by using the scaling relation $N \propto l^{-d_f}$, where $N$ is the number of boxes counted for a given box length $l$, to obtain the correlation fractal dimension $d_f$.}

{It is noteworthy that if, instead of merely counting the number of boxes containing at least one spin of the percolating cluster, we sum the total number of spins in this cluster within each box, we obtain the order parameter fractal dimension $d_l$.}

In Fig.\ \ref{figS4} we show the convergence of the instantaneous fractal dimension $D_f(T,t)$ as a function of Monte Carlo step $t$, at a fixed temperature $T$ in the 2d Ising model. The temperature  is  in units of $T_c$. We use a square lattice with lateral size $L=1024$.  We associate $D_f(T,t)$ with each measure obtained using our method to determine the fractal dimension. After reaching the equilibrium value $d_f(T)$
\begin{equation}
	\label{dT}
	d_f(T)=\frac{1}{t_f-t_i+1}\sum_{t=t_i}^{t_f} D_f(T,t),
\end{equation}
around which it fluctuates. The average values are taken from a step  $t_i$ at which it is thermalized, until a maximum number of monte calor steps $t_f$ is reached. Thus the final average $d_f(T)$ is taken over a number of realizations $N_r=L^2(t_f-t_i+1)$. 
The middle dashed line is the exact value $d_f(T_c)=7/4$ at the transition.  Upper curve $T=0.95$  and $d_f(0.95)=1.937(2)$ {(dashed green line)}; Lower curve $T=1.15$  and $d_f(1.15)=1.45(3)$ {(dashed black line)}. Note that the fluctuations above $T_c$ are larger than those below $T_c$.

We exhibit in the Fig.\ 1 of the main text $d_f(T)$ as a function of $T$.  We use a square lattice and the same procedure as Fig.\ \ref{figS4} above. 
Note that the time to flip the spin is much less than that to compute a fractal dimension, which is not done in CUDA. Consequently, due to the large time necessary to take the average Eq.\ (\ref{dT}) we use only a few points close to $T_c$. We check out our results  by imposing $d_f(Tc)=7/4$ and using $d_f(T) \approx f(T)=7/4+a(T-T_c)+b(T-T_c)^2$, to points close to the transition and fitting $f(T)$ to the data.

Finally, in Table \ref{Table2} we show the values of $T_c$ for some values of $L$ obtained using $f(T)$ above.  We exhibit the values for $L=1024 \times m$, for $m=1,2,...,6$. Precision depends on the size $L$ of the time $t_i$ that we expect to stabilize and the final time $t_f$. They define the number of realizations $N_r$. We see that for the values used in the simulations, we just see fluctuations around the value $T_c=1$. This shows that we have already reached the limit of precision. This supports our result $d_f=7/4$, the dashed line in Fig. 1 of the main text.

\begin{table}[h!]
	\centering
	\begin{tabular}{c | c c c c c c} 
		\hline
		\hline
		$L$ & $1024$ & $2048$ & $3072$ & $4096$ & $5120$ & $6144$ \\ [0.5ex] 
		\hline\hline
		$t_i$ & $85000 $  &   $110000$  & $50000$  & $79980$  & $79990$  & $80000$ \\ 
		$t_f$ & $87500 $  &   $110500$  & $50100$  & $80000$  & $80000$  & $80010$ \\
		$ 10^{-6} \times N_r$  & $2621 $ &   $2097$ & $372$     & $352$     & $288$  & $415$  \\
		$10^{4} \times\Delta T_c$  & $-8 \pm 6 $  &   $-0.4 \pm 3$ & $ 6 \pm 3$  & $-2 \pm 3$  & $0.7 \pm 3$  & $0.3 \pm 4$ \\ [1ex] 
		\hline
		\hline
	\end{tabular}
	\caption{Values of the difference   $ \Delta T_c = T_c-1 $  as function of $L$.  Values of $T_c$ taken from $d_f(T)=7/4+a(T-T_c)+b(T-T_c)^2$, close to transition. $N_r=L^2(t_f-t_i+1)$ is the number of realizations. Here $t_i$ is a time after stabilization, $t_f$ is the final execution time. Note that the deviations are of the same order as the error, in some cases smaller, therefore, they are just fluctuations within the precision limit.}
	\label{Table2}
\end{table}

\end{document}